\documentclass[aps,prl,twocolumn,reprint,superscriptaddress,nofootinbib,preprintnumbers,longbibliography]{revtex4-2}

\usepackage[english]{babel}
\usepackage[utf8]{inputenc}
\usepackage{amssymb}
\usepackage{amsmath}
\usepackage{color}
\usepackage{hyperref}
\usepackage{bbm}
\usepackage{epsfig}
\usepackage{multirow}

\usepackage[normalem]{ulem}

\usepackage[dvipsnames]{xcolor}
\usepackage{xspace}

\newcommand{\e}{\varepsilon}
\newcommand{\s}{\sigma}

\newcommand{\up}{\uparrow}
\newcommand{\down}{\downarrow}
\newcommand{\w}{\omega}

\newcommand{\avg}[1]{\langle {#1} \rangle }
\newcommand{\GF}[1]{\langle\!\langle #1\rangle\!\rangle}

\newcommand{\vk}{\mathbf{k}}

\newcommand{\JK}{J_{\rm K}}

\newcommand{\dk}{d^\dagger}
\newcommand{\ck}{c^\dagger}

\newcommand{\da}{d^{}}
\newcommand{\ca}{c^{}}
\newcommand{\spin}{\vec{s}}
\newcommand{\Spin}{\vec{S}}

\newcommand{\beq}{ \begin{equation} } 
\newcommand{\eeq}{ \end{equation} }
\newcommand{\beqa}{\begin{eqnarray}}
\newcommand{\eeqa}{\end{eqnarray}}
\newcommand{\nn}{\nonumber}

\newcommand{\ie}{\textit{i.e.}}

\newcommand{\logicsec}[1]{\color{black}\emph{#1}\color{black}.---}
\newcommand{\fig}[1]{Fig.~\ref{fig:#1}}
\newcommand{\figs}[1]{Figs.~\ref{fig:#1}}
\newcommand{\eq}[1]{Eq.~(\ref{#1})}

\hypersetup{
    bookmarks=false,        
    colorlinks=true,        
    linkcolor=red,        
    citecolor=blue,        
    filecolor=blue,      
    urlcolor=blue
}

\begin{document}

\title{Quantum spin liquid in an RKKY-coupled two-impurity Kondo system}

\author{Krzysztof P. W{\'o}jcik}
\email{kpwojcik@ifmpan.poznan.pl}
\affiliation{Physikalisches Institut, Universit\"{a}t Bonn, Nussallee 12, D-53115 Bonn, Germany}
\affiliation{Institute of Molecular Physics, Polish Academy of Sciences, 
			 Smoluchowskiego 17, 60-179 Pozna{\'n}, Poland}
\affiliation{Institute of Physics, Maria Curie-Sk\l{}odowska University, 20-031 Lublin, Poland}

\author{Johann Kroha}
\email{kroha@th.physik.uni-bonn.de}
\affiliation{Physikalisches Institut, Universit\"{a}t Bonn, Nussallee 12, D-53115 Bonn, Germany}

\date{\today}

\begin{abstract}
  We consider a 2-impurity Kondo system with spin-exchange coupling within
  the conduction band. Our numerical renormalization group calculations show
  that for strong intraband spin correlations the competition of these
  correlations with Kondo spin screening stabilizes a metallic spin-liquid
  phase of the localized spins without geometric frustration. For weak Kondo
  coupling the spin liquid and the Kondo singlet phase are separated by two
  quantum phase transitions and an intermediate RKKY spin-dimer phase, while 
  beyond a critical coupling they are connected by a crossover. The results 
  suggest how a quantum spin liquid may be realized in heavy-fermion systems 
  near a spin-density wave instability.
\end{abstract}

\maketitle

\logicsec{Introduction} 
Quantum spin liquids (QSLs) are systems of interacting spins with 
long-range entanglement, but without long-range magnetic 
order down to the lowest observed temperatures. 
The concept of non-local entanglement was first introduced in 1973 
by P. W. Anderson proposing the resonant valence-bond (RVB) state as the 
ground state of an antiferromagnetically coupled spin system on a triangular 
lattice \cite{Anderson1973},
which was later applied to the high-temperature cuprate superconductors 
\cite{Anderson1987}.
At the present day, QSLs are a wide and intensive 
research field in its own right \cite{Norman2016,Balents2017,Takagi2019}, 
due to the possibility of hosting fractional 
excitations and, thus, inducing new, unconventional quantum states of matter. 
Most theoretical studies are done on insulating and geometrically frustrated 
or topological spin lattice models, 
such as the triangular, kagome, next-nearest neighbor coupled or honeycomb
lattices \cite{Norman2016,Balents2017,Takagi2019,Sachdev1991}.

However, some important, possible realizations of QSLs, such as near a
magnetic heavy-fermion quantum phase transition (QPT) or in cuprate 
superconductors \cite{Anderson1987}, require metallic states and are in
general not geometrically frustrated. Such systems are generically described
by Anderson lattice models, i.e., localized magnetic impurities on a lattice
hybridizing with a sea of itinerant conduction electrons.
They often exhibit a QPT
\cite{Lohneysen2007,Paschen2021} between a paramagnetic 
heavy Fermi liquid induced by the Kondo effect \cite{Kondo1964,Hewson_book} and 
a magnetically ordered phase due to the Ruderman-Kittel-Kasuya-Yosida (RKKY)
spin-spin coupling $Y$ \cite{RK,K,Y} mediated by the conduction electrons.
A QSL in such systems must be stabilized, on the one hand, against the spin
extinction due to Kondo singlet formation and, on the other hand, against
magnetic ordering. Previous work on metallic, correlated systems predicted a 
QSL for strong spin exchange coupling between local and conduction electron 
spins, but was limited to mean-field treatments \cite{Andrei1989,Coleman1989}.
Some candidates for metallic QSLs in frustrated Kondo lattices 
have been proposed \cite{Nakatsuji2006Mar,Lucas2017Mar,Zhao2019Dec,Majumder2022May,Takabatake2022}. However, the existence and origin 
of metallic QSLs without geometric frustration have remained elusive. 

In the present work, we study a two-impurity
Anderson (2iA) model which incorporates the salient features of 
Anderson lattice systems, i.e., Kondo singlet formation, 
non-local RKKY interaction and spin correlations within the conduction 
band, and at the same time is amenable to numerically exact solution 
by the numerical renormalization group (NRG) \cite{WilsonNRG,NRG_RMP,fnrg}. 
Using NRG calculations we find, in addition to the QPT from a Kondo singlet to
a dimer singlet phase at weak RKKY coupling, 
another QPT at strong coupling from the dimer to a new phase 
characterized by fractional impurity spectral density and spin 
correlations -- a two-impurity realization of a QSL which is not driven by
geometric frustration, but by the competition of two screening effects. 
This phase is continuously connected to the Kondo singlet phase via strong 
Kondo spin exchange. That is, the Kondo effect stabilizes the QSL 
against RKKY dimer formation. We discuss experimental realizations of this QSL 
phase and its relevance for metallic lattice spin systems.

\begin{figure}[tb!]
\centering
\includegraphics[width=0.8\linewidth]{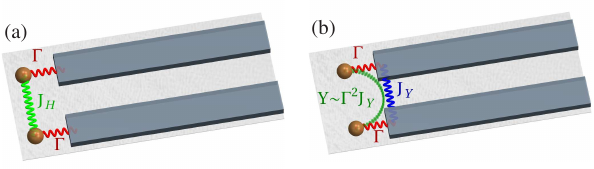}
\caption{Illustration of the 2iA model with (a) JV direct
  spin exchange $J_{\rm H}$ between the impurities and (b)
  conduction-electron-mediated (RKKY) coupling via Heisenberg exchange
  $J_Y$, \eq{H-2iA}.
  }
\label{fig:models}
\end{figure}

\logicsec{Model}
Early works by Jones and Varma (JV) \cite{Jones1,Jones2,Jones3}
effectively considered a two-impurity Kondo model where each spin-$1/2$
impurity is coupled to its own metallic host by the Kondo
coupling $J_{\rm K}$, and the RKKY interaction $Y$ is replaced by a direct
Heisenberg exchange $J_{\rm H}$ between the two impurities,
see \fig{models} (a). 
In this model, the Kondo singlet ground state and a dimer
singlet phase, characterized by $\pi/2$ or $0$ scattering phase shift of each
impurity, respectively, are separated by a QPT as a function of
the control parameter $J_{\rm H}/T_{\rm K}^0$, where $T_{\rm K}^0$ is the
Kondo temperature of a single Kondo impurity \cite{Hewson_book}. 
However, any particle-hole (PH) asymmetry of the JV model changes
the QPT into a crossover \cite{Fye1994,Affleck1995}.
Moreover, it has even been shown that with a
proper modeling of the RKKY interaction, the antiferromagnetic contribution
to $Y$ stems from the PH-asymmetric component of the effective Hamiltonian
\cite{Fabian2018}. Hence, a QPT does generically not occur in
two-impurity systems with one common host \cite{RKKYrange,MitchellBullaRKKY}, 
although the QPT may be restored by a counter-term compensating potential
scattering \cite{Fabian2018}, by suppressing charge transfer between
the screening channels \cite{Zarand2006Oct}, or by self-consistency 
in the auxiliary 2iA model in dynamical mean-field treatment of the lattice \cite{Gleis2022}. 
Another difficulty of the JV treatment \cite{Jones1,Jones2,Jones3} is that,
unlike $J_{\rm H}$, the true RKKY interaction is not independent of the
Kondo exchange $\JK$, but rather $Y \sim \JK^2$ \cite{RK,K,Y}.
This leads to a dynamical frustration effect and a universal suppression
of the Kondo scale depending on $Y$, $T_{\rm K}^0\to T_{\rm K}(Y)$, as has been shown
experimentally \cite{Bork2011} and theoretically
\cite{Hans2017}. These problems have called into question the relevance 
of the JV quantum critical point for QPTs in Kondo and Anderson lattice 
systems.

Therefore, we consider a maximally symmetric 
2iA model with conduction-electron-mediated 
impurity-spin coupling but without interhost potential scattering
as follows, cf.~\fig{models} (b),
\begin{eqnarray}
H_{\rm 2iA} &=&\sum_{\alpha\vk\s} \e_{\vk} \ck_{\alpha\vk\s} \ca_{\alpha\vk\s} 
	+ \sum_{\alpha\vk\s} V 
        (\ck_{\alpha\vk\s} \da_{\alpha\s} +\dk_{\alpha\s}\ca_{\alpha\vk\s} )
\nn\\
    &-&\frac{U}{2} \sum_{\alpha\s} n_{\alpha\s}
	+ U \sum_{\alpha} n_{\alpha\up} n_{\alpha\down} 
	+ J_Y \spin_1 \cdot \spin_2 .
\label{H-2iA}
\end{eqnarray}
Here, $\dk_{\alpha\s}$, $\da_{\alpha\s}$ are the operators for electrons with
spin $\s=\uparrow,\downarrow$ on impurity $\alpha=1,\,2$, 
$n_{\alpha\s}=\dk_{\alpha\s}\da_{\alpha\s}$, coupled to their respective
conducting leads with operators $\ck_{\alpha\vk\s}$, $\ca_{\alpha\vk\s}$ and 
dispersion $\e_{\vk}$ by the hybridization $V$, which we assume momentum 
independent (local) for simplicity. $U$ denotes the onsite repulsion 
on the impurity sites. Taking $-U/2$ for the impurity single-particle 
level and a flat conduction electron density of states $\rho$ at the Fermi 
level ensures PH symmetry. The hybridization generates the Kondo 
spin-exchange coupling $\JK=4|V|^2/U$ and the single-particle impurity-level
broadening $\varGamma=\pi\rho|V|^2$ \cite{Hewson_book}.  
The conduction electron spin at the impurity site 
of host $\alpha$ is defined in terms of the vector of Pauli matrices 
$\vec\sigma$ and $\ca_{\alpha\s} \equiv \sum_{\vk} \ca_{\alpha\vk\s}$ as 
\begin{eqnarray}
\spin_\alpha = \frac{1}{2}\sum_{\s\s'}\ck_{\alpha\s} 
\vec\sigma_{\s\s'} \ca_{\alpha\s'}\ . 
\label{cSpin}
\end{eqnarray}
The last term in the Hamiltonian (\ref{H-2iA}) describes
a Heisenberg exchange of strength $J_Y$ between these conduction electron 
spins. Thus, this model represents a single host as far as spin correlations
are concerned. It preserves PH symmetry, there is no 
symmetry-breaking charge transfer between the leads, and at the same time
it generates an RKKY coupling between the impurity 
spins $\Spin_{\alpha}$, mediated by the conducting hosts, 
$Y\,\Spin_1\cdot\Spin_2$, where $Y\approx(\rho \JK)^2\,J_{Y}/4$ and  $\Spin_{\alpha}$ 
is defined analogous to \eq{cSpin}.   
The latter turns out to be crucial for the phase diagram of the
system, while the asymmetry effects are discussed in Ref.~\cite{Wojcik2022b}. 

We analyze the system using the full density matrix 
approach to NRG \cite{Weichselbaum,AndersSchiller1}, 
which allows for precise calculation of thermal expectation values 
and determination of retarded Green functions $\GF{\ldots}^{\rm ret}$ 
directly in their Lehmann representation.
We use the open-access code \cite{fnrg}
as a basis for our programs, 
with discretization parameter $\Lambda=2.5$ and energy cutoff
at each iteration $6<E_{\rm cut}<6.5$;
see also Ref.~\cite{Wojcik2022b}.
The different phases of the system will be characterized by the
normalized local spectral density of impurity electrons, 
${\mathcal A}_T(\w)=-\varGamma \,{\rm Im} \GF{d^{}_{\alpha\s};d^{\dagger}_{\alpha\s}}^{\rm ret}(\w)$
and of conduction electrons at the impurity site,
${\mathcal B}_T(\w)=-2D\,{\rm Im}\GF{c^{}_{\alpha\s};c^\dagger_{\alpha\s}}^{\rm ret}(\w)$.
Equivalently, we will also use the corresponding, $T$-dependent 
differential conductances, 
$G(T)=-\int d\w f'(\w)A_T(\w)$ and 
$g(T)=-\int d\w f'(\w)B_T(\w)$, where $f'(\w)$ is the $\w$-derivative 
of the Fermi-Dirac distribution function.  
We define the single-impurity Kondo scale $T_{\rm K}^0$ as that 
temperature where $G(T)$ reaches 1/2 of its maximum fixed-point 
value $G_0$ during the NRG flow (see also below). 

\begin{figure*}[t!]
\includegraphics[width=0.9\linewidth]{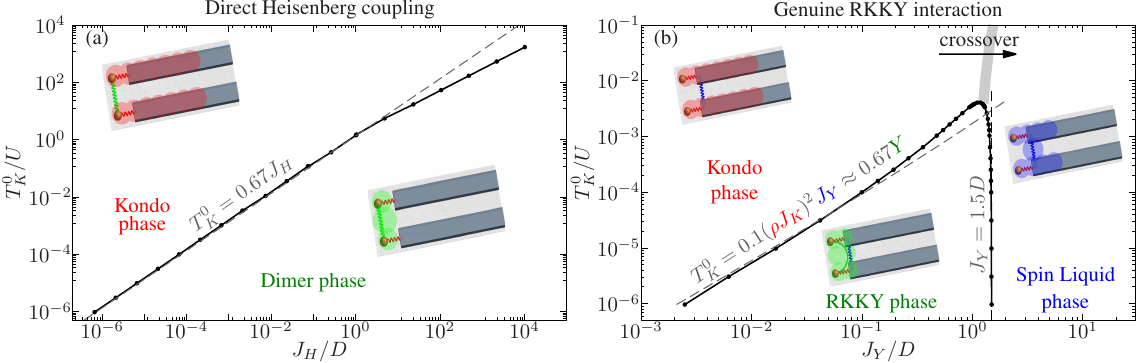}
\caption{
  Phase diagrams of the PH-symmetric 2iA model
  with (a) direct inter-impurity exchange $J_{\rm H}$ and (b) conduction-host
  mediated RKKY interaction $Y\approx (\rho \JK)^2J_Y/4$. The black dots 
  mark the QPT positions calculated by NRG, connected by lines for clarity. The dashed lines
  represent $J_{\rm H}=1.5\, T_{\rm K}^0$ and $T_{\rm K}^0=1.5\,Y$, respectively, as indicated. The
  insets illustrate the spatial structure of the spin correlations in the different
  phases. }
\label{fig:PDs}
\end{figure*}

\logicsec{Kondo vs.~Heisenberg quasiparticles}
In order to illu\-strate the competition between Kondo and intraband
screening we first recollect two limiting cases of \eq{H-2iA} separately.

(1)  $\JK>0$, $J_Y=0$: The single-impurity spin $1/2$ Kondo or Anderson model
is well understood \cite{Kondo1964,WilsonNRG,Hewson_book}. 
The local spin exchange $\JK$ between impurity and host 
induces the Kondo effect, \ie, the formation of a spatially extended,
many-body spin-singlet state comprised of the impurity spin and a multitude
of conduction electron states for $T<T_{\rm K}^0$, the Kondo screening cloud
\cite{SimonAffleckPRL,Borda}. 
This happens for arbitrarily small $\JK > 0$, since $\JK$ is subjected to the 
renormalization group flow toward the strong coupling fixed point
\cite{WilsonNRG}. 
This ground state is a Fermi liquid, and its excitations are characterized
by the Abrikosov-Suhl resonance in the impurity spectral density of 
unit height, $\mathcal{A}_0(0)=1$, and width $T_{\rm K}^0$, the effective bandwidth
of the Kondo quasiparticles. The conduction spectral density at the impurity 
site $\mathcal{B}_0(\w)$ is proportional to the inverse quasiparticle
lifetime and vanishes in a Fermi liquid manner as $\sim (\w/T_{\rm K}^0)^2$.

(2) $\JK=0$, $J_Y>0$: 
An analogous spin-screening occurs when two metallic leads are coupled 
to each other by a Heisenberg interaction $J_Y$ without impurities.
This interlead coupling leads to the destruction of free band electrons 
and formation of another Fermi liquid phase, signaled by the low-frequency
vanishing of the spectral density at the coupled sites as $\mathcal{B}_0(\w)\sim (\w/D)^2$.
However, unlike in the Kondo case, this happens only above a critical coupling
strength, $J_Y>J_Y^{**}$, since $J_Y$ is not renormalized to a strong coupling 
fixed point at low energies \cite{Hans2017}. Our NRG calculations show that 
$J_Y^{**} \approx 1.5D$ for metallic leads with 
rectangular, normalized density of states of width $2D$, 
$\rho(\w) = 1/(2D)$. 
The spin correlations induced by $J_Y>J_Y^{**}$ in the hosts form a
spatially extended object of size $\xi_H\approx v_F/J_Y$
(with $v_F$ the Fermi velocity), which we call the \emph{Heisenberg cloud}.

\logicsec{Direct Heisenberg exchange}
For comparison below, we now map out the phase diagram of the 2iA
model with direct Heisenberg exchange. This amounts to replacing 
the last term of the Hamiltonian (\ref{H-2iA}) with the direct impurity 
coupling term $J_{\rm H} \Spin_1\cdot \Spin_2$. 
As expected, we find for this PH-symmetric Anderson model the same 
QPT between a Kondo singlet and a dimer singlet phase as in the 
JV two-impurity Kondo model \cite{Jones1,Jones2,Jones3}. It is marked 
by discontinuous jumps of the impurity and host spectral densities 
$\mathcal{A}_{T=0}(0)$, $\mathcal{B}_{T=0}(0)$ from $1$ to $0$ and $0$ to
$1$, respectively, from the Kondo screened phase to the dimer phase. 
Thus, this constitutes a coupling-decoupling QPT in the charge sector
where the Kondo phase is governed by the Kondo quasiparticles described
above and the dimer phase by free Bloch electrons.
Nevertheless, the impurity and conduction spins remain correlated for
all $0 < J_{\rm H} < \infty$, and static correlations
$\avg{\vec{S}_1 \cdot \vec{S}_2}_{T=0}$ are continuous functions 
of $J_{\rm H}$ through the QPT \cite{Jones2}.
In \fig{PDs} (a) we show the 
resulting phase diagram in the $J_{\rm H}-T_{\rm K}^0$ plane near $T=0$, where $U=D/2$ 
was used throughout and $T_{\rm K}^0$ determined from the single-impurity NRG flow
for a given parameter set ($U$, $\varGamma$) as described above. In particular, we 
find the phase transition line as $J_{\rm H} = 1.5 T_{\rm K}^0$, with slight 
deviations for $J_{\rm H}/D\gg 1$, as compared to $J_{\rm H}=2.2 T_{\rm K}^0$ for the 
JV two-impurity Kondo model \cite{Jones2}.

\begin{figure}[btp]
\includegraphics[width=\linewidth]{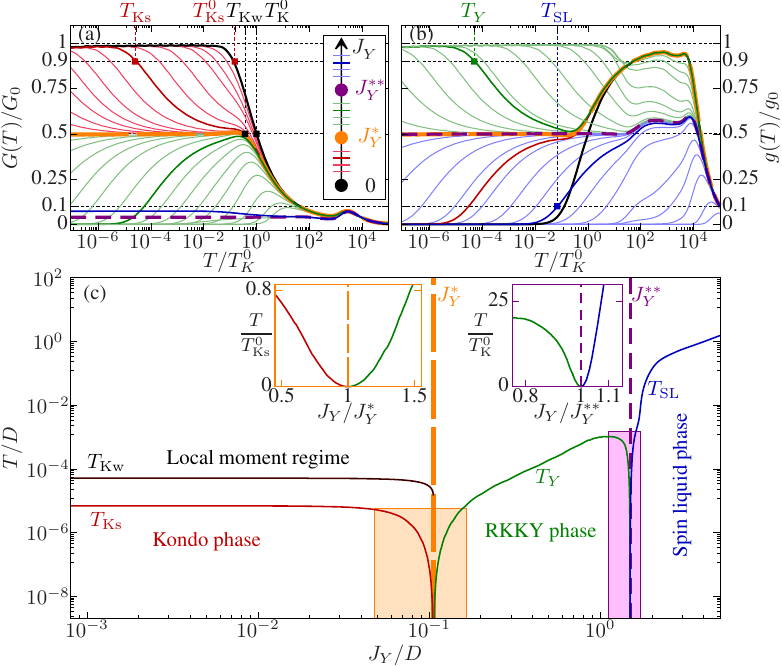}
\caption{
  RG flow and crossover scales in the Kondo, RKKY and QSL phases.
  (a) and (b) show the temperature-dependent impurity conductance,
  $G(T)$, and the host conductance at the site of the impurity,
  $g(T)$, respectively. The crossover scales are identified as
  explained in the text. The long (short) dashed lines were obtained
  for the quantum critical points at $J_Y^{}=J_Y^*$ ($J_Y^{}=J_Y^{**}$),
  respectively. (c) The crossover scales $T_{{\rm Kw}}$, $T_{{\rm Ks}}$,
  $T_{Y}$, and $T_{\rm SL}$ as a function of $J_Y$. The insets
  represent the zoomed-in, shaded areas around the QPTs. All
  computations were done for the parameter values $U/D=1/2$,
  $\varGamma/U=0.0488$, corresponding to $T_{\rm K}^0/U \approx 10^{-4}$.
}
\label{fig:RGflow}
\end{figure}

\logicsec{RKKY coupling and QSL}
We now consider the full 2iA model (\ref{H-2iA}). As can be
seen from the discussion of the Heisenberg cloud above, the
inter-host spin coupling $J_Y$ destabilizes not only the Kondo phase via the
RKKY interaction (see below), but also the (almost) free Bloch states in the host
towards an inter-host spin-correlated phase, when the coupling exceeds the
characteristic energy scale of the destabilized phase,
i.e., $Y\gtrsim T_{\rm K}^0$ and $J_Y\gtrsim D$, respectively.
We, therefore, extend our study to large $J_Y$ of the order of the 
conduction bandwidth $D$, using $U=D/2$ and $\varGamma=0.0488U$
corresponding to $T_{\rm K}^0\approx 10^{-4}U$, for the
numerical evaluations. The temperature dependence of the impurity
conductance $G(T)$ and of the host conductance at the impurity site
$g(T)$, each normalized to its unitary value, $G_0$ and $g_0$,
is shown in \figs{RGflow} (a), (b), respectively.
It is seen that there exist three stable, low-energy fixed points
characterized by the $T=0$ conductances as
\begin{eqnarray}
(1)&&\ {\rm Kondo:}\  G(0)=G_0, \ g(0)=0 \ \ {\rm (red\ curves)} \nn \\
(2)&&\ {\rm RKKY:}\  G(0)=0, \ g(0)=g_0 \ \ {\rm (green\ curves)}  \nn \\
(3)&&\ {\rm QSL:}\ \ \ \ G(0)={\rm fractional},\ g(0)=0 \ {\rm\ (blue\ curves)} \nn 
\end{eqnarray}
These are attained for different interhost coupling strengths $J_Y$ separated by
QPTs at $J_Y=J_Y^*$ (long-dashed curve) and $J_Y=J_Y^{**}$
(short-dashed curve) as shown in the figure. 
The behaviors in the phases (1) and (2) are as in the 2iA model with
direct Heisenberg exchange (see above). The phases (1) and (2) are, therefore,
identified with the well-known Kondo and the RKKY (dimer) phases, respectively.
The fixed point (3) is strikingly different. Because of its fractional impurity
quasiparticle spectral weight but Fermi-liquid-like low-energy host conductance,
$g(T)\sim T^2$, and together with the spin correlatons discussed below,
we identify this phase with a quantum spin liquid (QSL) state.

Each of the $G(T)$ and $g(T)$ curves in \figs{RGflow} (a), (b) is characterized
by two temperature scales, $T_{{\rm Kw}}(J_Y)$ at which the system flows away
from the high-$T$ free-local-moment fixed point, and a strong-coupling scale  
on which the respective Kondo, RKKY or QSL
strong-coupling fixed points are attained. From the NRG flow, in the
Kondo phase ($J_Y<J_Y^*$) we define $T_{{\rm Kw}}(J_Y)$ as the temperature where $G_T(0)$
reaches 1/2 of its maximum value $G_0$, and $T_{{\rm Ks}}(J_Y)$ as the temperature where
it becomes $G(T)>0.9G_0$, cf.~\fig{RGflow} (a).
While for $J_Y = 0$, $T_{{\rm Kw}}(0)$ coincides with the single-impurity Kondo scale
$T_{\rm K}^0$ and is proportional to the strong-coupling scale $T_{{\rm Ks}}(0)$,
$T_{{\rm Kw}}(J_Y)$ and $T_{{\rm Ks}}(J_Y)$ are in general independent, depending on
on two parameters $T_{\rm K}^0$ and $J_Y$. 
In the RKKY phase ($J_Y^*<J_Y<J_Y^{**}$),  and in the QSL phase ($J_Y>J_Y^{**}$),
the respective strong-coupling scales $T_Y)$ and $T_{\rm SL}$ are defined as the
temperature where $g(T)/g_0= 0.9$ and $g(T)/g_0= 0.1$, cf.~\fig{RGflow} (b).
The dependence of the crossover scales on the RKKY parameter $J_Y$ is shown
in \fig{RGflow} (c). 
We note that all strong-coupling scales vanish quadraticlly at the respective
quantum critical points; see insets of \fig{RGflow} (c). 
However, the weak-coupling Kondo scale $T_{{\rm Kw}}$ remains finite
with a suppression factor of $T_{{\rm Kw}}(J_Y^*)/T_{\rm K}^0 \approx 1/e$ (with
$e\approx 2.718$ is Euler's constant), and ceases to exist beyond $J_Y^*$,
in agreement with the analytic result of Ref.~\cite{Hans2017}.

\begin{figure}[t!]
\centering
\includegraphics[width=\linewidth]{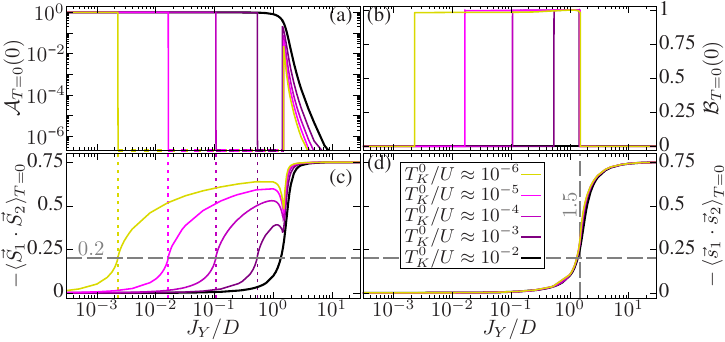}
\caption{Signatures of the QPTs in physical quantities, depending
  on the RKKY coupling parameter $J_Y$ at $T=0$.
  (a) and (b) show the normalized impurity and conduction spectral densities, ${\mathcal A}_0(0)$ and
  ${\mathcal B}_0(0)$, respectively.
  The static spin correlations between the impurity spins and between the conduction spins
  at the impurity site are shown in (c) and (d). 
		 }
\label{fig:PhysQ}
\end{figure}

The dependence of physical quanities on the RKKY parameter $J_Y$ are
shown in \fig{PhysQ} for fixed $T_{\rm K}^0$.
For sufficiently low $T_{\rm K}^0$, the quasiparticle
spectral densities ${\mathcal A}_0(0)=G(0)$ and ${\mathcal B}_0(0)=g(0)$ exhibit discontinuous
jumps signaling the two phase transitions at $J_Y^*$ and $J_Y^{**}$, respectively.
In the QSL phase ($J_Y>J_Y^{**}$), ${\mathcal A}_0(0)$ has fractional values
which decay to zero into the QSL phase. This signals incomplete
(and deep in the QSL phase vanishing) Kondo screening of the impurity spins. 
The static impurity spin correlations [\fig{PhysQ} (c)] behave continuously at the
Kondo-to-RKKY transition, as expected from a JV-like QPT, but exhibit a sharp cusp at the RKKY-to-QSL QPT and approach the singlet value of $-3/4$ deep in the QSL phase.
By contrast, the conduction spin correlations  [\fig{PhysQ} (d)]
show no indication of singular behavior. The fact that anomalous behavior appears
only in the impurity (fractional charge excitations and spin correlations)
but not in the conduction electron sector, gives rise to defining this phase as
a two-impurity analogon of a quantum spin liqid. Above a critical value of $T_{\rm K}^0$
all quantities behave continuously, \ie{}, there exists no QPT, see curves
for $T_{\rm K}^0\approx 10^{-2}U$ in
\fig{PhysQ}.

All these results can now be summarized in the complete phase diagram
of the RKKY 2iA model \eq{H-2iA} shown in \fig{PDs} (b). While the Kondo-to-RKKY QPT
is essentially identical to the one of the model with direct Heisenberg
exchange [\fig{PDs} (a)] with a critical line of $Y\approx 1.5 T_{\rm K}^0$,
the RKKY-to-QSL line is independent of $T_{\rm K}^0$ and occurs at a large
coupling of $J_Y\approx 1.5 D$. This is expected, because the
couplings $Y$ and $J_Y$ destabilize the quasiparticles
in the Kondo and in the RKKY phases at their
respective, characteristic energy scales, $T_{\rm K}^0$ and $D$.
The independence of the RKKY-to-QSL transition line of $T_{\rm K}^0$
implies that it must meet with the Kondo-to-RKKY transition line.
That is, the RKKY phase must terminate at a critical value
of $T_{\rm K}^0$, as seen in \fig{PDs} (b), and
the zero-temperature QSL and Kondo phases are continuously connected via
a crossover at large $T_{\rm K}^0$.

\logicsec{Conclusion and experimental realization}
We discovered a new type of quantum spin liquid in a model of two magnetic 
ions coupled to a metallic electron system with additional spin coupling 
play within the conduction band. This phase is characterized by 
non-local spin entanglement and fractional charge excitations on the 
magnetic ions, an analog of spin liquids in lattice systems, however without
the need for geometrical frustration. Instead, the quantum spin liquid is 
stabilized by a dynamical frustration effect, i.e., the interplay of 
Kondo spin screening and strong conduction-electron spin correlations 
with correlation energy of the order of the conduction bandwidth. 
In heavy-fermion compounds or Anderson lattice systems, such correlations 
may be achieved near a magnetic, e.g., spin-density wave instability 
within the conduction electron system. The new phase may also be 
realized experimentally in two-impurity systems with low conduction 
bandwidth, like magic-angle bilayer graphene \cite{Bistritzer2011}, 
with magnetic coupling between the layers.

\begin{acknowledgments}
Stimulating discussions with Frithjof Anders, Fabian Eickhoff, Andreas Gleis,
Mohsen Hafez-Torbati, and Kacper Wrze\'{s}niewski are gratefully acknowledged.
This project was financially supported by the Deutsche Forschungsgemeinschaft
(DFG, German Research Foundation) under Germany's Excellence Strategy –
Cluster of Excellence \textit{Matter and Light for Quantum Computing}, ML4Q
(390534769), and through the DFG Collaborative Research Center CRC 185 OSCAR
(277625399). K.P.W. acknowledges funding by the Alexander von Humboldt
Foundation and support from the Polish National Science Centre through grant no.~2018/29/B/ST3/00937. 
\end{acknowledgments}

%

\end{document}